\documentclass[aps,pre,showpacs,twocolumn]{revtex4}%
\usepackage{amsfonts}
\usepackage{amsmath}
\usepackage{amssymb}
\usepackage{graphicx}%
\providecommand{\U}[1]{\protect\rule{.1in}{.1in}}

\begin{document}
\title{Transport calculations in complex materials: A comparison of the Kubo formula,
the Kubo-Greenwood formula and the microscopic response method}
\author{M. -L. Zhang and D. A. Drabold}
\affiliation{Department of Physics and Astronomy, Ohio University, Athens Ohio 45701}

\begin{abstract}
Recently we have introduced the microscopic response method (MRM) to compute
the conductivity and Hall mobility for complex system with topological and
thermal disorder, which is more convenient than the Kubo formula. We prove
that for a canonical ensemble the MRM leads to the same expression as the Kubo
formula. When the gradient of carrier density is small, the MRM reduces to the
widely used Kubo-Greenwood formula.

\end{abstract}
\date[October 18, 2010]{}
\maketitle



Linear response theory\cite{ku57,kyn} is a rigorous and complicated procedure
used to compute transport coefficients. It constructs the observable
macroscopic response by averaging the operator of microscopic response over
the density matrix of system. Then, to obtain transport coefficients, one has
to calculate an imaginary time integral resulting from the commutator between
the density matrix and microscopic response\cite{kth}. If a system has several
types of elementary excitations and the interactions among elementary
excitations are strong, it is difficult to fulfil the imaginary time integral
with controllable approximations. For example, the imaginary time integral
prevents researchers from computing all the important contributions in
conductivity and Hall mobility of small
polarons\cite{lan62,fir,sch65,hol68,ald,lyo}. Amorphous semiconductors and
semiconducting polymers require approximations beyond small polarons: the
low-lying excited states often contain both localized states and extended
states\cite{motda}, and the electron-phonon interactions in localized states
are much stronger than those in extended states\cite{dra,tafn}. Even for the
lowest order self-consistent approximation\cite{4c}, it is difficult to apply
the Kubo formula, and consistently include all important contributions for
conductivity or Hall mobility.

To compute the transport coefficient for a mechanical perturbation, the
microscopic response method (MRM) is more convenient\cite{short,4c} than Kubo
formula. A mechanical perturbation such as the coupling with an external field
can be expressed via additional terms in the Hamiltonian\cite{ku57,kyn}, and
the wave function $\Psi^{\prime}$ of system in an external field at a later
moment is determined by its initial value and the time-dependent
Schr\"{o}dinger equation. The microscopic response can be expressed in terms
of the changes in wave function induced by the external field\cite{short,4c}.
The ensemble average and coarse-grained average needed to compute the
macroscopic response (transport coefficient) can then be carried out at the
final stage. Thus for a mechanical perturbation, we are able to avoid the
imaginary time integral in the Kubo formula \cite{short,4c}.

The Kubo-Greenwood formula (KGF)\cite{Gre} has been implemented in many
\textit{ab initio} codes to calculate the dielectric function and AC
conductivity. However the KGF is based on a simplified expression for the
current density, which is borrowed from the kinetic theory of gas.

In this paper, we prove that for mechanical perturbations, the MRM is
equivalent to the Kubo formula. To fulfil this aim, we first write out the
observable macroscopic current density to first and second order in external
field in the Kubo formulation.Then the same procedures are carried out with
the MRM\cite{short}. We see that the macroscopic response calculated in the
two methods are the same. We discuss the connection between density matrix and
transition amplitudes at different order of perturbation. We will see why the
MRM allows easy classification of transport processes and expression for
transport coefficient compared to the Kubo formulation\cite{ku57}. We show
that the current density used by Greenwood is justified only when the gradient
of carrier density is small.

In this work, we use the Schr\"{o}dinger picture. Consider a system with $N$
electrons and $\mathcal{N}$ nuclei in an electromagnetic field with potentials
$(\mathbf{A},\phi)$, at time $t$, the many-electron state of system is
described by $\Psi^{\prime}(\mathbf{r}_{1},\mathbf{r}_{2},\cdots
,\mathbf{r}_{N};t)$. To save space, we will not write out the nuclear
coordinates explicitly. $\Psi^{\prime}$ satisfies the Schr\"{o}dinger equation%
\begin{equation}
i\hbar\partial\Psi^{\prime}/\partial t=H^{\prime}\Psi^{\prime},\text{
}H^{\prime}=H+V(t), \label{tds1}%
\end{equation}
where $V(t)$ is the interaction between the system and external field. The
time dependence in $V(t^{\prime})$ comes from the external field. $H$ is the
Hamiltonian of the system without external field. We use $|m\rangle$ or
$\Psi_{m}$ and $E_{m}$ to denote the $m^{th}$ stationary state and the
corresponding eigenvalue of the $N$ electrons + $\mathcal{N}$ nuclei system:
$H|m\rangle=E_{m}|m\rangle$. If the system is in a thermal bath at temperature
$T$, then before introducing $V(t)$, the equilibrium density operator is%
\begin{equation}
\widehat{\rho}=\sum_{m}|m\rangle P_{m}\langle m|\text{, }P_{m}=e^{-\beta
E_{m}}/Z\text{,} \label{eqd}%
\end{equation}
where $Z=\sum_{n}e^{-\beta E_{n}}$ is the partition function. In an external
electromagnetic field, the velocity operator $\mathbf{v}_{i}$ for the $i^{th}$
particle is\cite{lv3}%
\begin{equation}
\mathbf{v}_{i}=m^{-1}[\mathbf{P}_{i}-e\mathbf{A}(\mathbf{r}_{i};t)],
\label{vel}%
\end{equation}
where $\mathbf{r}_{i}$ and $\mathbf{P}_{i}=-i\hbar\nabla_{\mathbf{r}_{i}}$ are
the position and momentum operators of the $i^{th}$ particle, and $e$ is the
charge of electron. Because velocity and position cannot be simultaneously
measured, one has to symmetrize the velocity and position operators in the
current density operator. One may conjecture that\ the current density
operator at point $\mathbf{r}$ is\cite{mah}:%
\begin{equation}
\widehat{\mathbf{j}}(\mathbf{r})=\frac{e}{2}\sum_{i=1}^{N}[\mathbf{v}%
_{i}\delta(\mathbf{r}-\mathbf{r}_{i})+\delta(\mathbf{r}-\mathbf{r}%
_{i})\mathbf{v}_{i}]. \label{co}%
\end{equation}

In the MRM\cite{kubo,short,4c}, we avoided $\widehat{\mathbf{j}}(\mathbf{r})$.
Now we show that Eq.(\ref{co}) leads to a proper microscopic current
density\cite{short,4c}. Because a mechanical perturbation can be expressed
with additional terms in Hamiltonian, the states at time $t$ can be described
by a wave function which is determined by the initial conditions. The
microscopic current density at time $t$ and point $\mathbf{r}$ in state
$\Psi^{\prime}(\mathbf{r}_{1},\mathbf{r}_{2},\cdots,\mathbf{r}_{N};t)$ is%
\begin{equation}
\mathbf{j}_{m}(\mathbf{r};t)=\int d\tau\Psi^{\prime\ast}\widehat{\mathbf{j}%
}(\mathbf{r})\Psi^{\prime}, \label{defc}%
\end{equation}
where $d\tau=d\mathbf{r}_{1}d\tau^{\prime}$, and $d\tau^{\prime}%
=d\mathbf{r}_{2}\cdots d\mathbf{r}_{N}$. Integrating by parts, one has%
\begin{equation}
\mathbf{j}_{m}(\mathbf{r};t)=-\frac{e^{2}N}{m}\mathbf{A}(\mathbf{r})\int
d\tau^{\prime}\Psi^{\prime\ast}\nabla_{\mathbf{r}}\Psi^{\prime} \label{avcm}%
\end{equation}%
\[
+\frac{i\hbar eN}{2m}\int d\tau^{\prime}(\Psi^{\prime}\nabla_{\mathbf{r}}%
\Psi^{\prime\ast}-\Psi^{\prime\ast}\nabla_{\mathbf{r}}\Psi^{\prime}),
\]
where the arguments of $\Psi^{\prime}$ in Eq.(\ref{avcm}) are $(\mathbf{r}%
,\mathbf{r}_{2},\cdots,\mathbf{r}_{N};t)$. Eq.(\ref{avcm}) has been
independently derived from the principle of virtual work\cite{kubo}, the
continuity equation\cite{short} and the polarization density\cite{4c}. The
current operator given in Eq.(\ref{co}) is correct, and it will bridge the
Kubo formulation and the MRM.

To write out the macroscopic response in the Kubo formula, we notice that the
time evolution for a system involving mixed states is included in the density
matrix. The basis set should be a group of wave functions without any time
dependence\cite{tol}. In the $|m\rangle$ representation, the matrix elements
of $\widehat{\mathbf{j}}(\mathbf{r})$ are%

\[
\langle n|\widehat{\mathbf{j}}(\mathbf{r})|m\rangle=\frac{Nei\hbar}{2m}\int
d\tau^{\prime}[\Psi_{m}\nabla_{\mathbf{r}}\Psi_{n}^{\ast}-\Psi_{n}^{\ast
}\nabla_{\mathbf{r}}\Psi_{m}]
\]%
\begin{equation}
-\frac{Ne^{2}}{m}\mathbf{A}(\mathbf{r})\int d\tau^{\prime}\Psi_{n}^{\ast}%
\Psi_{m}, \label{bed2}%
\end{equation}
where the arguments of $\Psi_{m}$ and $\Psi_{n}$ are $(\mathbf{r}%
,\mathbf{r}_{2},\cdots,\mathbf{r}_{N})$. With $\rho$ in Eq.(\ref{eqd}) as the
initial condition, one can use perturbation theory to solve the Liouville
equation to any order in $V(t)$. The density matrix at time $t$ is
$\rho^{\prime}(t)=\rho+\rho^{(1)}(t)+\rho^{(2)}(t)+\cdots$. To first order in
$V(t)$, the deviation $\rho^{(1)}(t)$ from $\rho$ is\cite{kth}%
\begin{equation}
\langle m|\rho^{(1)}(t)|n\rangle=\frac{1}{i\hbar}\int_{-\infty}^{t}dt^{\prime
}e^{i(t-t^{\prime})(E_{n}-E_{m})/\hbar} \label{1vd}%
\end{equation}%
\[
\langle m|V(t^{\prime})|n\rangle(P_{n}-P_{m}).
\]
The conductivity can be read off from the macroscopic current density:%
\begin{equation}
\mathbf{j}^{(1)}(\mathbf{r},t)=\sum_{mn}\langle m|\rho^{(1)}(t)|n\rangle
\langle n|\widehat{\mathbf{j}}(\mathbf{r})|m\rangle. \label{1c}%
\end{equation}
To 2nd order in $V(t)$, the deviation $\rho^{(2)}(t)$ is%
\begin{equation}
\langle m|\rho^{(2)}(t)|n\rangle=-\frac{1}{\hbar^{2}}e^{it(E_{n}-E_{m})/\hbar
}\sum_{k}\{ \label{2vd}%
\end{equation}%
\[
(P_{n}\int_{-\infty}^{t}dt^{\prime}\int_{-\infty}^{t^{\prime}}dt^{\prime
\prime}+P_{m}\int_{-\infty}^{t}dt^{\prime\prime}\int_{-\infty}^{t^{\prime
\prime}}dt^{\prime})
\]%
\[
e^{it^{\prime\prime}(E_{k}-E_{n})/\hbar}e^{it^{\prime}(E_{m}-E_{k})/\hbar
}\langle m|V(t^{\prime})|k\rangle\langle k|V(t^{\prime\prime})|n\rangle
\]%
\[
-(\int_{-\infty}^{t}dt^{\prime}\int_{-\infty}^{t^{\prime}}dt^{\prime\prime
}+\int_{-\infty}^{t}dt^{\prime\prime}\int_{-\infty}^{t^{\prime\prime}%
}dt^{\prime})
\]%
\[
e^{it^{\prime\prime}(E_{k}-E_{n})/\hbar}e^{it^{\prime}(E_{m}-E_{k})/\hbar
}\langle m|V(t^{\prime})|k\rangle P_{k}\langle k|V(t^{\prime\prime}%
)|n\rangle\},
\]
The 2nd order macroscopic response $\mathbf{j}^{(2)}(\mathbf{r},t)$ is
obtained from Eq.(\ref{1c}) by replacing $\langle m|\rho^{(1)}(t)|n\rangle$
with $\langle m|\rho^{(2)}(t)|n\rangle$. We are going to compare
Eqs.(\ref{1vd},\ref{1c},\ref{2vd}) and $\mathbf{j}^{(2)}(\mathbf{r},t)$ with
the corresponding quantities in the MRM.

In the MRM, the macroscopic response is given by \cite{short,4c}:%
\begin{equation}
\mathbf{j}(\mathbf{r},t)=\sum_{n}P_{n}\langle\Psi_{n}^{\prime}(t)|\widehat
{\mathbf{j}}(\mathbf{r})|\Psi_{n}^{\prime}(t)\rangle. \label{tdsj}%
\end{equation}
If the initial state is $\Psi_{n}$, then the state $\Psi_{n}^{\prime}(t)$ of a
system at time $t$ in an external field can be determined\cite{tol} by
applying perturbation theory to Eq.(\ref{tds1}):%
\begin{equation}
\Psi_{n}^{\prime}(t)=a^{(0)}(n,t)\Psi_{n} \label{tdw}%
\end{equation}%
\[
+\sum_{m}a^{(1)}(mn,t)\Psi_{m}+\sum_{m}a^{(2)}(mn,t)\Psi_{m},
\]
where%
\begin{equation}
a^{(0)}(n,t)=e^{-iE_{n}t/\hbar}, \label{0t}%
\end{equation}
and%
\begin{equation}
a^{(1)}(mn,t)=-\frac{i}{\hbar}e^{-iE_{m}t/\hbar}\int_{-\infty}^{t}dt^{\prime}
\label{1t}%
\end{equation}%
\[
e^{i(E_{m}-E_{n})t^{\prime}/\hbar}\langle m|V(t^{\prime})|n\rangle,
\]
and%
\[
a^{(2)}(mn,t)=-\frac{1}{\hbar^{2}}e^{-iE_{m}t/\hbar}\sum_{k}\int_{-\infty}%
^{t}dt^{\prime}e^{i(E_{m}-E_{k})t^{\prime}/\hbar}%
\]%
\begin{equation}
\langle m|V(t^{\prime})|k\rangle\int_{-\infty}^{t^{\prime}}dt^{\prime\prime
}e^{i(E_{k}-E_{n})t^{\prime\prime}/\hbar}\langle k|V(t^{\prime\prime
})|n\rangle. \label{2t}%
\end{equation}

To first order in $V(t)$, the macroscopic current density is
\begin{equation}
\mathbf{j}^{(1)}(\mathbf{r},t)=\sum_{n}P_{n}[\langle\Psi_{n}^{(0)}%
(t)|\widehat{\mathbf{j}}(\mathbf{r})|\Psi_{n}^{(1)}(t)\rangle\label{1sj}%
\end{equation}%
\[
+\langle\Psi_{n}^{(1)}(t)|\widehat{\mathbf{j}}(\mathbf{r})|\Psi_{n}%
^{(0)}(t)\rangle].
\]
Substituting Eqs.(\ref{tdw},\ref{0t},\ref{1t}) into Eq.(\ref{1sj}), and using
the fact that $V(t^{\prime})$ and $\widehat{\mathbf{j}}(\mathbf{r})$ are
Hermitian operators, one finds the same result as Eqs.(\ref{1vd},\ref{1c}). To
second order in $V(t)$, the macroscopic current density is%
\begin{equation}
\mathbf{j}^{(2)}(\mathbf{r},t)=\sum_{n}P_{n}[\langle\Psi_{n}^{(0)}%
(t)|\widehat{\mathbf{j}}(\mathbf{r})|\Psi_{n}^{(2)}(t)\rangle\label{2sj}%
\end{equation}%
\[
+\langle\Psi_{n}^{(2)}(t)|\widehat{\mathbf{j}}(\mathbf{r})|\Psi_{n}%
^{(0)}(t)\rangle+\langle\Psi_{n}^{(1)}(t)|\widehat{\mathbf{j}}(\mathbf{r}%
)|\Psi_{n}^{(1)}(t)\rangle].
\]
Substituting Eqs.(\ref{tdw},\ref{0t},\ref{1t},\ref{2t}) into Eq.(\ref{2sj}),
the first term of Eq.(\ref{2sj}) is the same as the 1st term of $\mathbf{j}%
^{(2)}$ from Eq.(\ref{2vd}), the 2nd term of Eq.(\ref{2sj}) is the same as the
2nd term of $\mathbf{j}^{(2)}$ resulted from Eq.(\ref{2vd}). One can see that
the 3rd term Eq.(\ref{2sj}) equals the sum of the 3rd term and the 4th term of
$\mathbf{j}^{(2)}$ from Eq.(\ref{2vd}), if one notices: (1) three integrands
are the same; (2) the 3rd term in Eq.(\ref{2sj}) is a two-dimensional integral
in domain $[-\infty,t;-\infty,t]$; (3) the 3rd term of $\mathbf{j}^{(2)}$ is a
successive integration $\int_{-\infty}^{t}dt^{\prime}\int_{-\infty}%
^{t^{\prime}}dt^{\prime\prime}$; (4) the 4th term of $\mathbf{j}^{(2)}$ is a
successive integration $\int_{-\infty}^{t}dt^{\prime\prime}\int_{-\infty
}^{t^{\prime\prime}}dt^{\prime}$. The procedure is easy to proceed to any
order in field. The equation (\ref{tdsj}) does not use any specific property
of electromagnetic field, the procedure works for any mechanical perturbation.
Introducing current density operator (\ref{co}) is the key for the proof. In
the original MRM, one does not need current density operator, the macroscopic
response is obtained by averaging over the microscopic response (\ref{avcm})
over canonical distribution. Eqs.(\ref{co},\ref{tdsj}) established a
connection between two methods.

It is worthwhile to find the connection between the probability\ amplitudes in
Eqs.(\ref{0t}-\ref{2t}) and the density matrices in Eqs.(\ref{1vd},\ref{2vd}).
The element of the density matrix is the average the product of two
probability amplitudes over the $\mathcal{M}$ members in an ensemble\cite{tol}%
:%
\begin{equation}
\rho_{mn}=\frac{1}{\mathcal{M}}\sum_{\alpha}a_{\alpha}^{\ast}(n,t)a_{\alpha
}(m,t), \label{dev}%
\end{equation}
where $\alpha$ is the index of a member in the canonical ensemble. To the
first order in $V(t)$,%
\begin{equation}
\rho_{mn}^{(1)}=\frac{1}{N}\sum_{\alpha}a_{\alpha}^{\ast(0)}(n,t)a_{\alpha
}^{(1)}(m,t) \label{1r}%
\end{equation}%
\[
+\frac{1}{N}\sum_{\alpha}a_{\alpha}^{\ast(1)}(n,t)a_{\alpha}^{(0)}(m,t),
\]
where $a_{\alpha}^{(0)}(n,t)$ is the zero order transition amplitude from
initial state $|n\rangle$ to final state $|n\rangle$, $a_{\alpha}^{(1)}(m,t)$
is the first order transition amplitude from initial state $|n\rangle$ to
final state $|m\rangle$, $a^{(0)}(m;t)$ is the zero order transition amplitude
from initial state $|m\rangle$ to final state $|m\rangle$, $a^{(1)}(n,t)$ is
the first order transition amplitude from initial state $|m\rangle$ to final
state $|n\rangle$. With these explanations, substituting Eqs.(\ref{0t}%
,\ref{1t}) into Eq.(\ref{1r}), one reaches Eq.(\ref{1vd}), which was obtained
from Liouville equation. To 2nd order in $V(t)$,%
\begin{equation}
\rho_{mn}^{(2)}=\frac{1}{\mathcal{M}}\sum_{\alpha}\{a_{\alpha}^{\ast
(0)}(n,t)a_{\alpha}^{(2)}(m,t) \label{r2}%
\end{equation}%
\[
+a_{\alpha}^{\ast(2)}(n,t)a_{\alpha}^{(0)}(m,t)+a_{\alpha}^{\ast
(1)}(n,t)a_{\alpha}^{(1)}(m,t)\}.
\]
In the 1st term of (\ref{r2}), the initial state is $|n\rangle$, $a_{\alpha
}^{(2)}(m,t)$ is the 2nd order transition amplitude through intermediate
states $|k\rangle$. By means of Eqs.(\ref{0t},\ref{2t}), the first term of
Eq.(\ref{r2}) is the same as the 1st term of Eq.(\ref{2vd}). In the second
term of (\ref{r2}), the initial state is $|m\rangle$, $a_{\alpha}^{(2)}(n,t)$
is the 2nd order transition amplitude through intermediate states $|k\rangle$
to final state $|n\rangle$. The second term of Eq.(\ref{r2}) is the same as
the second term of Eq.(\ref{2vd}). In the 3rd term of (\ref{r2}), two final
states $|n\rangle$ and $|m\rangle$ come from a common initial state
$|k\rangle$, all states $\{|k\rangle\}$ satisfy $k\neq n$ and $k\neq m$ can be
taken as the initial state. In terms of the same trick in comparing the 3rd
term in Eq.(\ref{2sj}) and the sum of the 3rd and 4th terms of $\mathbf{j}%
^{(2)}(\mathbf{r},t)$ derived from Eq.(\ref{2vd}), we can see that the 3rd
term of (\ref{r2}) is the same as the sum of the 3rd term and the 4th term in
Eq.(\ref{2vd}).

We explain why the MRM is simpler than the Kubo formula for mechanical
perturbations. To a given order in residual interactions, various transport
processes contribute to a specific transport coefficient. In the MRM, each
transport process is composed of several elementary transitions caused by
external field and by residual interactions\cite{epjb,4c}. Because the
microscopic response is expressed by the wave function of system in external
field rather than density matrix, each elementary transition appears as a
transition amplitude\cite{4c}. According to Eq.(\ref{tdw}), the state at $t$
is a linear superposition of the various order changes induced by the external
field. By means of Eqs.(\ref{avcm},\ref{1sj},\ref{2sj}), gradient operator
connects two components of the final state\cite{4c}. In addition, the
transition amplitude of a higher order transition is constructed by first
making a product of the sequence of first order amplitudes of elementary
transitions and then summing over all intermediate states. We can depict each
transport process with a diagram, which has one line connecting two components
of final state, and several other lines presenting elementary transitions. To
a given order of residual interactions, the topology of diagrams can help us
classify and construct all possible transport processes\cite{4c}. In the Kubo
formulation, all time-dependence is included in density matrix, cf.
Eqs.(\ref{1c},\ref{bed2}). To a given order in external field, the change in
density matrix involves different members of the ensemble, cf. Eqs.(\ref{1r}%
,\ref{r2}). Besides, the density matrix is bilinear in transition amplitude.
Therefore for a transport process with more than one elementary transitions,
one cannot express it as a product of propagators.

Greenwood derived his conductivity expression from\cite{Gre}
\begin{equation}
\mathbf{j}(t)=eTr\{\rho^{\prime}(t)\mathbf{v}\}\mathbf{,} \label{gren}%
\end{equation}
where $\rho^{\prime}(t)$ is the density matrix of system in external field. We
show that Eq.(\ref{gren}) is justified only when the fluctuation in the
spatial distribution of carriers is small. Using Eq.(\ref{vel}) and the
commutation relation between $\mathbf{r}_{i}$ and $\mathbf{P}_{i}$,
Eq.(\ref{co}) becomes%
\begin{equation}
\widehat{\mathbf{j}}(\mathbf{r})=e\sum_{i=1}^{N}\delta(\mathbf{r}%
-\mathbf{r}_{i})\mathbf{v}_{i}-\frac{i\hbar e}{2m}\sum_{i=1}^{N}%
[\nabla_{\mathbf{r}_{i}}\delta(\mathbf{r}-\mathbf{r}_{i})]. \label{co1}%
\end{equation}
Averaging Eq.(\ref{co1}) over state $\Psi^{\prime}(\mathbf{r}_{1}%
,\mathbf{r}_{2},\cdots,\mathbf{r}_{N};t)$, Eq.(\ref{defc}) gives another
expression for microscopic current density:%
\begin{equation}
\mathbf{j}(\mathbf{r};t)=\frac{i\hbar e}{2m}\mathbf{\nabla}_{\mathbf{r}%
}n^{\prime}(\mathbf{r}) \label{trc}%
\end{equation}%
\[
+eN\int d\tau^{\prime}\Psi^{\prime\ast}m^{-1}[-i\hbar\nabla_{\mathbf{r}%
}-e\mathbf{A}(\mathbf{r})]\Psi^{\prime},
\]
where the arguments of $\Psi^{\prime}$ in Eq.(\ref{trc}) are $(\mathbf{r}%
,\mathbf{r}_{2},\cdots,\mathbf{r}_{N};t)$, $n^{\prime}(\mathbf{r})=N\int
d\tau^{\prime}\Psi^{\prime\ast}\nabla_{\mathbf{r}}\Psi^{\prime}$ is the number
density of electrons at point $\mathbf{r}$ in external field $(\mathbf{A}%
,\phi)$. The 1st term of Eq.(\ref{trc}) can be neglected only when the
gradient of carrier density is small. Using the corresponding relation between
the MRM and the Kubo formulation, the 2nd term of Eq.(\ref{trc}) reduces to
Eq.(\ref{gren}). It is obvious that Eq.(\ref{gren}) and the consequent KGF are
not suitable to the localized carriers in amorphous semiconductors and the $d$
and $f$ electrons in strong correlated systems. For a given error $\Delta x$
of position$,$ the error of velocity is $\Delta v\thicksim m^{-1}\hbar/\Delta
x$. The error of current density is $n^{\prime}e\Delta v\thicksim m^{-1}\hbar
n^{\prime}/\Delta x\thicksim m^{-1}\hbar\nabla_{\mathbf{r}}n^{\prime}$. If the
charge density is uniform like nearly electron gas, the uncertainty $\Delta x$
of position is infinity, the momentum of electron is completely determined
$\Delta p_{x}=0$. One can use kinetic expression for current density
$\mathbf{j}=2e\sum_{\mathbf{k}}\mathbf{v}_{\mathbf{k}}f_{\mathbf{k}}$, where
$f_{\mathbf{k}}$ is the distribution function.

In summary, we proved that for a mechanical perturbation the microscopic
response method is equivalent to and simpler than Kubo formula. To compute
transport coefficients for mechanical perturbations, the microscopic response
method is advantageous because of the ease of obtaining expression to a given
order of residual interactions consistently. When the gradient of carrier
density is small, the strict current density Eq.(\ref{trc}) reduces to the
kinetic expression (\ref{gren}).

We thank the Army Research Office for support under MURI W91NF-06-2-0026, and
the National Science Foundation for support under grants DMR 0903225.

\end{document}